\documentclass[a4paper,12pt]{article}
\usepackage{hyperref}
\usepackage{soul}
\usepackage[usenames,dvipsnames]{color}
\usepackage{amsmath, amssymb, slashed, epsf, color, graphicx}
\usepackage{epsfig}
\usepackage{ulem}
\def\Lc{{\cal L}}

\def\ba{\bar{a}}
\def\Wc{{\cal W}}
\def\Tc{{\cal T}}
\def\bz{{\bar{z}}}
\def\t{{\tau}}
\def\bt{{\bar{\tau}}}
\def\Gc{{\cal G}}
\def\Jc{{\cal J}}
\def\l{{\lambda}}

\def\bA{\bar{A}}
\date{}
\begin{document}
\begin{flushright}
 HRI/ST$1407$
\end{flushright}

\begin{center}

{\large\bf Phase Structure of Higher Spin Black Holes}\\[15mm]
Abhishek Chowdhury \footnote {E-mail: abhishek@hri.res.in},
Arunabha Saha \footnote{E-mail: arunabha@hri.res.in}\\
{\em Harish-Chandra Research Institute\\
Chhatnag Road, Jhusi, Allahabad - 211 019, India\\[20mm]} 
\end{center}
 \abstract{
 We revisit the study of the phase structure of higher spin black holes carried out  in arXiv$:1210.0284$ using the ``canonical formalism''. In particular we study the low as well as the high temperature regimes. We show that the Hawking-Page transition takes place in the low temperature regime. The thermodynamically favoured phase changes from conical surplus to black holes and then again to conical surplus as we increase temperature. We then show that in the high temperature regime the diagonal embedding gives the appropriate description. We also give a map between the parameters of the theory near the IR and UV fixed points with the effect that the ``good'' solutions near IR fixed point map to the ``bad'' solutions near the UV fixed point and vice versa.
 %We also give the map between solutions near the IR and UV fixed points. This ~We show the region in parameter space where Hawking-Page transition takes place at low temperature regime and how the phase structure %changes from conical surplus to black hole and back to conical surplus as we increase the temperature. In the process we arrive at an ``extremal thermal AdS '' branch. We then show that in the high temperature %regime the diagonal embedding solution gives the valid description and we also give a map between good and bad solutions near the UV and IR fixed points.   
}

\newpage
\section{Introduction}
Higher spin theories in various dimensions have been the object of interest for quite sometime now \cite{Fronsdal:1978rb,Fradkin:1987ks,Fradkin:1986qy,Vasiliev:1990tk,Vasiliev:1992av,Vasiliev:2003ev}. They have become a very useful arena for studying the nature of AdS/CFT dualities. In \cite{Klebanov:2002ja} and \cite{Sezgin:2002rt,Sezgin:2003pt}, the first example of a higher spin AdS/CFT duality was given. There it was conjectured that a theory of O(N) vector model in $2+1$ dimensions is dual to the higher spin theories in $AdS_4$. To be more precise the singlet sector of O(N) vector models was shown to be dual to the Vasiliev system with only even spins turned on. But in general these theories have a spectrum consisting of one copy of each spin ranging from $2$ to $\infty$ \cite{Vasiliev:2003ev}. 
\\\indent In 3 dimensions the complexity reduces quite a bit due to the fact that there are no bulk propagating degrees of freedom and due to the related fact that the spectrum can be truncated to any finite maximal spin N. With these simplifications the theories in 3 dimensions serve as good toy models to understand various aspects of both the higher spin theories and the AdS/CFT dualities. Higher spin theory in $3$ space-time dimensions was studied in \cite{Vasiliev:1995dn,Blencowe:1988gj}. In the latter the $SL(2,R)\times SL(2,R)$ Chern-Simons formulation for gravity in AdS space was extended to $SL(N,R)\times SL(N,R)$ theory and it was shown that the spectrum is that of fields of spin ranging from $2$ to N. The classical asymptotic symmetry algebra of higher spin theories in $AdS_3$ has been shown to match with W-symmetry algebra in \cite{Campoleoni:2010zq,Henneaux:2010xg}. See also \cite{Gaberdiel:2011wb,Campoleoni:2011hg} for further analysis of the asymptotic symmetry algebra. The 
first indication that the symmetry is present in the quantum regime was provided in \cite{Gaberdiel:2010ar}, where the one loop partition function in the bulk was calculated and shown to be equal to the vacuum character of the W-symmetry enhanced CFTs. Based on this, a duality between higher spin theories in $AdS_3$ and CFT with W-symmetry was proposed in \cite{Gaberdiel:2010pz}. Further elaborations of the proposal were done \cite{Gaberdiel:2011zw,Castro:2011iw, Gaberdiel:2012ku,Perlmutter:2012ds} and for a review see \cite{Gaberdiel:2012uj}. The topologically massive versions of these higher spin theories and their possible dualities to logarithmic CFTs was shown in \cite{Chen:2011vp,Bagchi:2011vr,Bagchi:2011td,Chen:2011yx}. 

%-------------------------------------------------------------------------------------------------

\subsection{Higher Spin Black Holes in $3$ dimensions}
\indent  In $3$ dimensions, the topology of a space-time with asymptotic AdS geometry is that of a solid torus. The contractible cycle is either spatial or temporal depending on whether we are in a thermal AdS background or a black hole background. In black holes the spatial  non-contractible cycle points towards the existence of a ``horizon''. For Euclidean black holes the temperature is defined by assuming a periodic time cycle. The periodicity is such that the horizon has no conical singularities i.e. the horizon is smooth. This periodicity in time  cycle is related to the inverse of the temperature of the black hole. 
\\ \indent In higher spin theories the concept of a metric is blurred by the fact that there are higher spin gauge transformations under which the metric is not invariant.$~$Hence, the normal procedure of identifying black hole geometries to metrics with horizons doesn't work. In \cite{Gutperle:2011kf}, a procedure to identify the higher spin black hole geometry in $AdS_3$, in the Chern-Simons formulation was given. There the black hole geometry was identified with those configurations where the connection is smooth in the interior of the torus geometry with a contractible temporal cycle. This is equivalent to demanding a trivial holonomy for the connection along the temporal cycle (i.e. it falls in the centre of the gauge group). This ensures that when the contractible cycle is shrunk to zero, the connection comes back to itself after moving around the cycle once. But this does not ensure that the corresponding metric will look like that of an ordinary black hole. In \cite{Ammon:2011nk} a gauge 
transformation was found in which the metric obtained resembled that of a conventional black hole. It was also shown that the RG flow by an irrelevant deformation triggered by a chemical potential corresponding to a spin $3$ operator takes us from the principal embedding of $sl(2,R)$ to the diagonal embedding of $sl(2,R)$ in $sl(3,R)$. Now, to get a higher spin black black hole a chemical potential corresponding to the independent charges had to be added so that the system is stable thermodynamically. So, a black hole solution with higher spin charges necessarily causes the system to flow from one fixed point to another. In \cite{Kraus:2011ds} the partition function for the black hole solution was obtained as a series expansion in spin $3$ chemical potential with $hs[\l]\times hs[\l]$ algebra (this gives the higher spin symmetry algebra when the spin is not truncated to any finite value) and matched with the known CFT 
results for free bosonic ($\l=1$) and free fermionic case ($\l=0$). This answer also matches the one for general $\l$ obtained from CFT calculations in \cite{Gaberdiel:2012yb}. A review of these aspects of black holes in higher spin theories can be found in \cite{Ammon:2012wc}.The $\lambda\rightarrow\infty$ limit for partition function was studied in \cite{Beccaria:2013dua}, where an exact expression for partition function and spin $4$ charge was obtained for any temperature and spin $3$ chemical potential. Analysis of a HS black holes in presence of spin $4$ chemical potential was done in \cite{Beccaria:2013gaa}. 
\\ \indent A different approach to study the thermodynamics of these black holes was carried out in \cite{Banados:2012ue,deBoer:2013gz,Perez:2012cf,Perez:2013xi}. A good variational principle was obtained by adding proper boundary terms to Chern Simons theories on manifolds with boundaries. The free energy was obtained from the on-shell action and an expression for entropy was obtained from that. This expression for entropy was different from that obtained in \cite{Gutperle:2011kf}. It was also shown that the stress energy tensor obtained from the variational principle mixes the holomorphic and antiholomorphic components of the connection. This formalism for obtaining the thermodynamics variables is referred to as the ``canonical formalism'' in the literature. The CFT calculations done in \cite{Gaberdiel:2012yb} seem to match with the ``holomorphic formalism'' given in \cite{Gutperle:2011kf}, but the canonical approach seems to be much more physically plausible. In \cite{Compere:2013nba} a possible solution 
to this discrepancy was suggested,where they changed the bulk to boundary dictionary in a way suited to the addition of chemical potential which deforms the theory. 
\\\indent In \cite{Li:2013rsa}, the process of adding chemical potential was unified for the full family of solutions obtained by modular transformation from the conical defect solution. The black holes that we talked about is only one member of the family. It was shown that the same boundary terms need to be added to the action to get a good variational principle for all members of the family. There the definitions for all thermodynamic quantities for any arbitrary member of the family were obtained.

%--------------------------------------------------------------------------------------------------------------------------------

\subsection{Phase structure of higher spin black holes in $AdS_3$}
The phase structure of spin $3$ black holes in $AdS_3$ was studied in \cite{David:2012iu} using the holomorphic variables. In the principal embedding of $sl(2,R)$ in $sl(3,R)$ (with spectrum consisting of fields with spin $2$ and $3$), they found 4 solutions to the equations corresponding to a trivial holonomy along the time circle. They allowed for a non-zero spin $3$ charge even when the corresponding chemical potential is taken to zero. It was shown there that of the $4$ branches one is unphysical as its entropy is negative. Of the remaining three branches one is the BTZ branch (here spin $3$ charge goes to zero as chemical potential goes to $0$), one is the extremal branch (having a non-zero charge configuration at zero temperature or chemical potential) and a third branch. The negative specific heat of the extremal branch makes it an unstable branch. A more analytical treatment of phase structure was done in \cite{Chen:2012ba} for spin $3$ and $\tilde{4}$ black holes.
\\\indent The phase diagram given there shows that the BTZ and extremal branch exist only in the low temperature regime, after which the thermodynamic quantities for this two branches do not remain real. The third branch that is present has real thermodynamic variables at all temperatures. It is shown there that at low temperatures the thermodynamics quantities have the correct scaling behaviour with temperature (from the point of view of a possible dual CFT description) only for the BTZ branch. The third branch, which exists for all temperatures, does not have the correct scaling behaviour for thermodynamic quantities at very high temperature. At low enough temperatures where the BTZ and extremal black hole solution exist, the BTZ branch has the lowest free energy followed by the extremal branch and then the third branch. But at higher temperatures the third branch is the only surviving branch, i.e. the only branch with real thermodynamic quantities.
\\\indent They then argued that the correct thermodynamics at high temperature is given by the diagonal embedding. The spectrum in the diagonal embedding has a pair of fields with spin $\frac{3}{2}$, a pair with spin $1$ and a spin $2$ field \cite{Campoleoni:2011hg}. The diagonal embedding can be thought of as the starting point of a RG flow initiated by the spin $\frac{3}{2}$ chemical potential and ending in the principal embedding. The temperature scaling behavior the thermodynamic quantities are found to be correct at high temperature. Near the zero chemical potential (for spin $\frac{3}{2}$) limit, i.e. near the starting point of the RG flow, the holonomy equations have 2 real solutions; among them the one with a lower free energy was conjectured to be the ``third branch''(from the principal embedding) at the end of the RG flow. This was done with the assumption that the two solutions which survive at all temperature in the principal embedding match with these two real solutions at the UV fixed point. 
The "third branch" has the correct scaling behaviour (w.r.t. the thermodynamic quantities in diagonal embedding) and it was argued that beyond the point where BTZ and extremal branches 
of the principal embedding cease to exist, this third branch takes over and it is actually the black hole solution in the diagonal embedding.
\\\indent In summary they showed that the principal embedding is the correct IR picture valid at low temperature regime and diagonal embedding is the correct UV picture valid in the  high temperature regime.

%----------------------------------------------------------------------------------------------------------

\subsection{Our Work} 
In this work we study the phase structure of $SL(3,R)\times SL(3,R)$ higher spin system in the canonical formalism. We will first work in the principal embedding. We will be using the definitions of thermodynamic quantities for conical surplus solution (which go to the thermal AdS branch when chemical potential and spin $3$ charges are taken to zero) given in \cite{Li:2013rsa}. The conical surplus has a contractible spatial cycle.~So, we demand that the holonomy of connection along this cycle be trivial.~Using this condition we are able to get the undeformed spin $2$ and $3$ charges in terms of temperature and chemical potential for spin $3$ charge. We use this to study the phase structure of the conical surplus. From the phase diagram we see that there are 2 branches of solutions with real values for undeformed spin $2$ and spin $3$ charges for a given temperature and chemical potential. One of the branch reduces to the thermal AdS branch  (with zero spin $3$ charge) when the 
chemical potential is taken to zero. The other one is a new 
branch which like the extremal black holes has a non-trivial charge configuration even when chemical potential and/or temperature is taken to be zero. This we call the ``extremal thermal AdS'' branch. This extremal branch has a lower free energy for all values of $\mu$ and T.
\\\indent We then move to studying the phase structure of black hole in this embedding. We again solve for the charges in terms of chemical potential (m) and temperature (T), but now with the time cycle contractible. Here we get $4$ branches of solutions. We find that two of these branches have negative entropy and hence are unphysical. Among the other two branches, one is the BTZ branch (which reduces to BTZ black hole when chemical potential $m\rightarrow 0$) and the other is the extremal branch (having a non-trivial charge configuration for zero chemical potential and/or temperature). The extremal branch has negative specific heat and hence is unstable. The BTZ branch is stable, has lower free energy and hence is the dominant of the two good solutions. Given a chemical potential both the black hole and the thermal AdS solutions exist till a certain temperature, which is different for the black hole and thermal AdS. Crossing the respective temperatures 
leads to complex values of the thermodynamic quantities. 
\\\indent We then undertake a study of the phase structure for the conical surplus and black hole together. Between the $3$ solutions- the BTZ black hole, the extremal black hole and the thermal AdS branch we study which branch has the minimum free energy for a given chemical potential and temperature. We notice that for a particular chemical potential, at very low temperature the thermal AdS has the lowest free energy and then as we gradually increase the temperature the BTZ branch starts dominating over the thermal AdS.~This is the analogue of Hawking-page transition. After a particular temperature the extremal black hole also dominates over the thermal AdS though it is sub-dominant to the BTZ branch. Increasing the temperature further, the black holes cease to exist and the thermal AdS is the only solution. So far, we have not considered the extremal thermal AdS branch which has the lowest free energy of all the branches. If this branch is not absent (due to some 
physical reasons that we are unaware 
of) it will be the thermodynamically dominant branch all through the low temperature regime.~We will comment further about this branch later in the paper. 
\\\indent Next part of our study involves studying black holes in the diagonal embedding of $sl(2,R)$ in $sl(3,R)$. There is a consistent truncation where the spin $\frac{3}{2}$ fields are put to zero \cite{Castro:2011fm}. But here we don't want to do this. The reason being that we want to use the fact that this diagonal embedding is actually the UV limit of the flow initiated in principal embedding by the spin $3$ chemical potential. We want to study the full theory obtained from this procedure and there all the mentioned fields are present. First of all taking cue from the map given in \cite{David:2012iu} and we will be able to give a map between the parameters that we use at UV and IR fixed points. Also, here we obtain $4$ solutions to the holonomy equations and by similar arguments as above two of them are unphysical. Of the other two branches the one with the lower free energy is throughout stable. We also showed that the good solution near IR fixed point actually maps to the bad solution near 
the UV fixed point and vice 
versa. We give a plausible reasoning 
for this mapping between the good and bad branches.

%---------------------------------------------------------------------------------------------------------------

\subsection{Organization of the paper}
In section \ref{review} we give a brief review of the geometry of higher spin theories and their thermodynamics. In the section \ref{principal} we give the analysis for the thermodynamics of conical surplus, black hole and Hawking-Page transition for principal embedding. In section \ref{diagonal} we give a similar description for black hole in the diagonal embedding. Lastly, we give a summary of our results in section \ref{summary} and some possible directions for future studies in \ref{future}.

%-----------------------------------------------------------------------------------------------------------------

\section{Review of higher spin geometry in $AdS_3$ and thermodynamics}\label{review}
Let us briefly elaborate on the 'Canonical formalism' for BTZ Black Holes in higher spin scenarios. We will mostly follow the conventions given in \cite{deBoer:2013gz, Li:2013rsa}. In $ 2+1 $ dimensions higher spin theories coupled to gravity with negative cosmological constant can be written as a Chern-Simons theory with gauge group $ G \simeq SL(N,R) \times SL(N,R) $ \cite{Blencowe:1988gj}. For  $ N=2$ it reduces to ordinary gravity but for $ N \ge 3 $ depending on possible embeddings of the $ sl(2,R)$ subalgebra into $sl(N,R)$ it generates a spectrum of fields with different spins.
%For example the so called `principal embedding' corresponds to bulk fields of spin 2 ( the metric ) plus a tower of higher spin fields ranging from $ 3 ,\ldots, N $.
We are mostly interested in  an Euclidean Chern-Simons theory on a three-dimensional manifold $ M $ with the topology $S^1 \times D $ where the $S^1$ factor is associated with the compactified time direction and $\partial D \simeq S^1$. 
It is customary to introduce coordinates $(\rho, z, \bar{z})$ on $ M $, where $ \rho $ is the radial coordinate and $ \rho \rightarrow \infty $ is the boundary with the topology of a torus where the $z,\;\bar{z}$ coordinates are identified as $ z(\bar{z}) \simeq z(\bar{z}) + 2 \pi \simeq z(\bar{z})+ 2 \pi \tau(\bar{\tau}) $. For Chern-Simons theory the field strength is zero, so the connection is pure gauge. We will be working in a gauge where the connections have a radial dependence given by 
$$ A = b^{-1} db + b^{-1} a b  \qquad \bar{A} = b db^{-1} + b\bar{a}b ^{-1} $$ 
with $ b= b(\rho)=e^{\rho L_0}$ and $a, \bar{a}$ being functions of boundary $z,\;\bar{z}$ coordinates only.\\ The holonomies associated with the identification along the temporal direction are 
\begin{equation}\label{timeholonomy}
Hol_{\tau, \bar{\tau}}(A) = b^{-1} e^h b \qquad Hol_{\tau, \bar{\tau}}(\bar{A}) = b e^{\bar{h}} b^{-1}
\end{equation}
where the matrices $ h $ and $ \bar{h} $ are
\begin{equation}
h = 2 \pi (\tau a_z + \bar{\tau} a_{\bar{z}}) \qquad \bar{h} = 2 \pi (\tau \bar{a}_z + \bar{\tau}\bar{a}_{\bar{z}})
\end{equation}
Triviality of the holonomy forces it to be an element of the center of the gauge group and a particularly interesting choice which corresponds to the choice for uncharged BTZ black hole gives 
\begin{equation}\label{holonomy condition}
Tr [h\cdot h ] = - 8\pi^2 \qquad Tr[h\cdot h \cdot h] =0 
\end{equation}
A different choice of the center element is synonymous to a scaling of $\tau$ and hence is not very important for us as we focus on a particular member of the centre of the group and are not interested in a comparative study between various members.  

 With this setup in mind the Euclidean action is
 $$ I^{(E)} = I^{(E)}_{CS} +  I^{(E)}_{Bdy}$$
 where
 $$ I^{(E)}_{CS} = CS[A]-CS[\bA], \quad CS[A]=\frac{i k_{cs}}{4 \pi} \int_M Tr[A\wedge dA+\frac{2}{3}A\wedge A\wedge A ]  $$
For a good variational principal on the manifold we need to add some boundary terms to the above action. To get a variation of action of the form $\delta I\sim Q_i \delta \mu_i$ (for grand canonical ensemble) we need to add a boundary term of the form 
 $$  I^{(E)}_{Bdy} = - \frac{k_{cs}}{2 \pi} \int_{\partial M} d^2 z\; Tr\left [(a_z - 2 L_1)a_{\bar{z}} \right ] - \frac{k_{cs}}{2 \pi} \int_{\partial M} d^2 z\; Tr\left [(\ba_{\bar{z}} - 2 L_{-1})a_{z} \right ] $$
\\\indent We will be interested in an asymptotically AdS boundary which will give rise to the $W_N$ algebra as the asymptotic symmetry algebra in the absence of any chemical potential. This is satisfied by the connections written in the Drinfeld-Sokolov form
 \begin{align}\label{DScondition}
 a &= \left ( L_1 + Q \right ) dz - \left ( M + \ldots \right ) d\bar{z} \\
 \bar{a} &= \left ( L_{-1} - \bar{Q} \right )d\bar{z} + \left ( \bar{M} + \ldots \right )dz
 \end{align}
with $ \left [ L_{-1}, Q \right ] = \left [ L_1, M \right ]=0 $   (and similarly for $\bar{Q}, \; \bar{M}$).
 We adopt a convention that the highest (lowest) weights in $a_z \,(\bar{a}_{\bar{z}})$  are linear in the charges, and the highest (lowest) weights in $\bar{a}_z \, (a_{\bar{z}})$ are linear in the  chemical potentials corresponding to charges other than spin $2$. The convention for definition of chemical potential that we use is given by
 \begin{align}\label{chemical potential}
 Tr \left [ (a_z - L_1)(\bar{\tau}-\tau)a_{\bar{z}} \right ] &= \sum_{i=3}^{N} \mu_i Q_i \\
 Tr \left [ (-\bar{a}_{\bar{z}} + L_{-1}) (\bar{\tau}-\tau)\bar{a}_z \right ] &=  \sum_{i=3}^{N} \bar{\mu}_i \bar{Q}_i
 \end{align}
 Varying $ I^{(E)} $ on-shell we arrive at
 \begin{equation}\label{onshellvariation}
 \begin{split}  
 \delta I^{(E)}_{os} = -\ln Z =& -2\pi i k_{cs} \int_{\partial M} \frac{d^2 z}{4 \pi^2 Im(\tau)} \;Tr  \bigg [ (a_z - L_1) \delta \left ((\bar{\tau} - \tau) a_{\bar{z}} \right ) + \left ( \frac{a^2_z}{2} + a_z a_{\bar{z}} - \frac{\bar{a}^2_z}{2} \right ) \delta \tau   \\
 &  \qquad \qquad \qquad \qquad \quad \;\;\; - (-\bar{a}_{\bar{z}} + L_{-1}) \delta \left ((\bar{\tau} - \tau) \bar{a}_{z} \right ) - \left ( \frac{\bar{a}^{2}_{\bar{z}}}{2} + \bar{a}_{\bar{z}} \bar{a}_{z} - \frac{a^2_{\bar{z}}}{2} \right ) \delta \bar{\tau} \bigg]\\
 =& - 2\pi i k_{cs} \int_{\partial M}  \frac{d^2 z}{4 \pi^2 Im(\tau)} \; \left ( T \delta \tau - \bar{T} \delta \bar{\tau} + \sum_{i=3}^{N} \left ( Q_i\delta \mu_i - \bar{Q}_i \delta \bar{\mu}_i \right ) \right )  
\end{split}
\end{equation}
 So, the added boundary terms are the correct one as we get the desired variation of the action on-shell.
\\\indent The black hole geometry that we discussed can be obtained by a $SL(2,Z)$ modular transformation acting on a conical surplus geometry and vice versa. This property was used in \cite{Li:2013rsa} to show that the variational principle for either geometry (or for that matter any geometry obtained by a $SL(2,Z)$ transformation on the conical surplus geometry) goes through correctly if we use the boundary terms given above. This in principle means that we have the same definition of stress tensor for all the members of the '$SL(2,Z)$' family and is given by 
\begin{equation}\label{stressall}
 \Tc=Tr[\frac{a_z^2}{2}+a_z a_{\bz}-\frac{\ba_z^2}{2}],\quad \bar{\Tc}=Tr[\frac{\ba^2}{2}+\ba_{\bz}\ba_z-\frac{a_{\bz}^2}{2}]
\end{equation}
The on-shell action evaluated for a member gives the free energy for that particular member. Please note that in arriving at equation (\ref{onshellvariation}) we have to go through an intermediate coordinate transformation pushing the $\tau$ dependence of the periodicity $z\simeq z+2\pi\tau $ to the integrand. This is necessary to make sure that the variation does not affect the limits of integration. However, this procedure of making periodicities of the coordinates constant is dependent upon the member of interest in the '$SL(2,Z)$' family. The free energy for any arbitrary member was evaluated in \cite{Li:2013rsa} and is given by
\begin{eqnarray}\label{freeall}
-\beta F&=& -I_{on-shell}\nonumber\\
&=&\pi i k_{cs}Tr\bigg[\left(h_Ah_B-\bar{h}_A\bar{h}_B\right)-2i(a_z-2L_1)a_{\bz}-2i(\ba_{\bz}-2L_{-1})\ba_z\bigg] 
\end{eqnarray}
where $h_A$ and $h_B$ are respectively the holonomy along the contractible and non-contractible cycles. 
%Note that since $\tau$ acts as a source for the stress tensor it is useful to  consider the following coordinate change 
% \begin{equation}
% z= \frac{1- i \tau}{2} w + \frac{1+ i \tau}{2} \bar{w}
% \end{equation}
% so that the identifications now become $ w(\bar{w}) \simeq w(\bar{w}) +2 \pi \simeq w(\bar{w}) + 2 \pi i (-i) $ and the modular parameter ($\tau$) explicitly appears in the boundary metric.\\
% \indent For constant connections $ a$, $\bar{a}$ solutions, we can use Vol$(\partial M ) = 4 \pi^{2} Im(\tau)$ and explicitly evaluate the Euclidean action on-shell with the boundary terms, obtaining the free energy F as 
% \begin{equation}
% \begin{split}
% -\beta F = \ln z = -2 \pi i k_{cs} Tr \bigg [ & \left ( \frac{a^2_z}{2} + a_z a_{\bar{z}} - \frac{\bar{a}^2_z}{2} \right ) \tau - \left ( \frac{\bar{a}^{2}_{\bar{z}}}{2} + \bar{a}_{\bar{z}} \bar{a}_{z} - \frac{a^2_{\bar{z}}}{2} \right ) \bar{\tau} \\
% & + (\bar{\tau} - \tau)(L_1 a_{\bar{z}} + L_{-1} \bar{a}_{z}) \bigg ]
% \end{split}
% \end{equation}
\\\indent Performing a Legendre transform of the free energy (i.e. from a function of chemical potentials/sources to function of charges) we arrive at an expression for the entropy. The expression for entropy of the black hole solution turns out to be 
\begin{equation}\label{entropy}
S = - 2 \pi i k_{cs} Tr \bigg [(a_z + a_{\bar{z}}) (\tau a_z + \bar{\tau} a_{\bar{z}}) - (\bar{a}_z + \bar{a}_{\bar{z}}) ( \tau \bar{a}_z + \bar{\tau} \bar{a}_{\bar{z}} )\bigg ]
\end{equation}
The conical surplus solutions are obtained by adding a chemical potential to thermal AdS like solutions. So, it is unexpected that they will suddenly develop properties reminiscent of objects with "horizon" (like having non-zero entropy) under presence of a small deformations. The calculation of entropy using the above method supports this intuition as we get zero entropy indeed for these solutions . %{\color{blue} This can also be understood from the point of view of a putative dual CFT. The states dual to the conical surplus are obtained by acting on the vacuum with a heavy operator and the vacuum being  a pure state this is also a pure state and hence has zero entropy.}
\\\indent All the above statements can very easily be generalized to non-principal embedding. The things that will be different are the value of the label 'k' associated with different $sl(2,R)$ embedding in $sl(3,R)$  and the definition of charges and chemical potentials. The value of k is related to $k_{cs}$ by
\begin{equation}
 k_{cs}=\frac{k}{2 Tr\left[\Lambda^0\Lambda^0\right]}, 
\end{equation}
where $k_{cs}$ is the label associated with the $SL(3,R)$ CS theory. The central charge of the theory for a particular embedding is given by $c=6k$. $\Lambda_{-1},\Lambda_0,\Lambda_1$ are the generators giving rise to the $sl(2,R)$ sub-algebra in the particular embedding. 

% \indent We do a similar analysis but with the canonical variables of \cite{deBoer:2013gz} and study the phase diagrams. We provide reasons why the principal embedding is valid in the low temperature regime and diagonal embedding is valid description at the high temperature without resorting to any scaling arguments. This supports the claim that the principal embedding is the IR description and the diagonal embedding is the UV description, though the central charge decreases from IR to UV. 
% \\ \indent In \cite{Li:2013rsa} it was shown that the boundary terms to be used are the same for all configurations obtained by $SL(2,Z)$ modular transformation matrix acting on black hole solutions . So, this works even for the conical surplus which is related to black hole by $\t_{\tiny CS}=-\frac{1}{\t_{BH}}$. We used this to study the regions in parameter space where the free energy of black hole solution dominates over the conical surplus, i.e. where the Hawking-Page transition takes place. We also study the phase diagram for both the principal and diagonal embedding of $SL(2)$ in $SL(3)$. 

\section{The principal embedding for sl(3,R)}\label{principal}
Here, we give the conventions for connections and the thermodynamic quantities that we use in our paper here. The connection that we use here are based on ~\cite{Gutperle:2011kf,Li:2013rsa,deBoer:2013gz}. We will confine our connections in the radial gauge and use the conventions for generators of $SL(3,R)$ group given in\footnote{Here we redefine our variables to absorb the k appearing in the connections given in \cite{deBoer:2013gz}} \cite{Ammon:2012wc}
\begin{eqnarray}\label{connection}
 a&=&(L_1-2\pi \Lc L_{-1}-\frac{\pi}{2} \Wc W_{-2}) dz+\frac{m T}{2}(W_2+4\pi\Wc L_{-1}-4\pi \Lc W_0+4\pi^2\Lc^2W_{-2})d\bz,\nonumber\\
 \ba&=&(L_{-1}-2\pi \bar{\Lc} L_1-\frac{\pi}{2}\bar{\Wc} W_2)d\bz+\frac{\bar{m} T}{2}(W_{-2}+4\pi\bar{\Wc}L_{-1}-4\pi\bar{\Lc}W_0+4\pi^2\bar{\Lc}^2W_2)dz.\nonumber\\
\end{eqnarray}
 We are interested in studying only non-rotating solutions, hence we require $g_{zz}=g_{\bz\bz}$ for the metric which when converted to the language of connections in radial gauge becomes
\begin{equation}\label{nonrot}
 Tr[a_{\bz}a_{\bz}-2a_{\bz}\ba_{\bz}+\ba_{\bz}\ba_{\bz}]=Tr[a_z a_z-2a_z\ba_z+\ba_z\ba_z].
\end{equation}
With our convention this is satisfied if $\bar{m}=-m$, $\bar{\Wc}=-\Wc$, $\bar{\Lc}=\Lc$
\\\indent These connections automatically satisfy the equations of motion $\left[ a_z,a_{\bz} \right] =0$. With our conventions the equation (\ref{chemical potential}) becomes  
\begin{equation}\label{chempotdef}
 Tr[(a_z-L_1)a_{\bz}(\bt-\t)]= 4 i m \pi \Wc.
\end{equation}
\\\indent So, demanding that $\Wc$, which is the measure of spin $3$ charge in our conventions be real, the chemical potential $\mu_3$ is imaginary, whose measure is given by  \textit{`im'}.

%------------------------------------------------------------------------------------------------------------------------------

\subsection{The Conical Surplus Solution}
\begin{figure}[h]
\epsfig{file=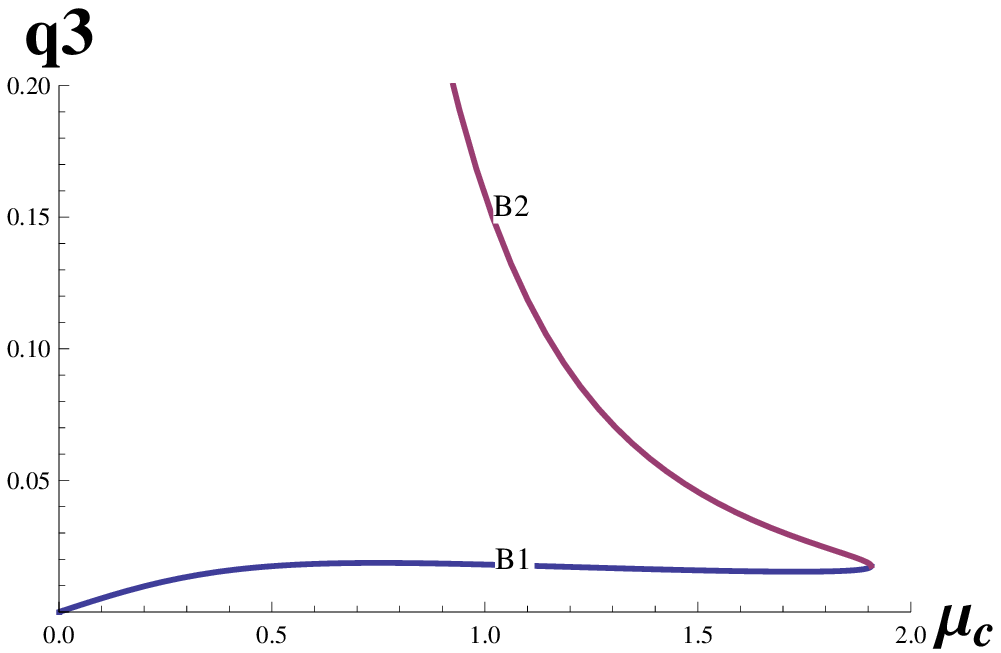, width =2in,height=2in}
\epsfig{file=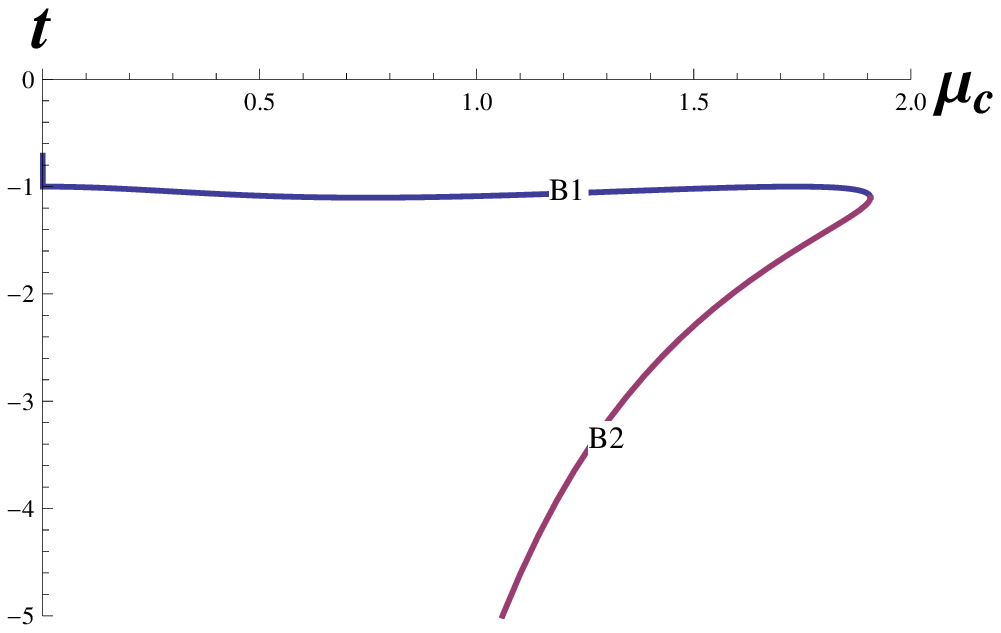, width =2in,height=2in}
\epsfig{file=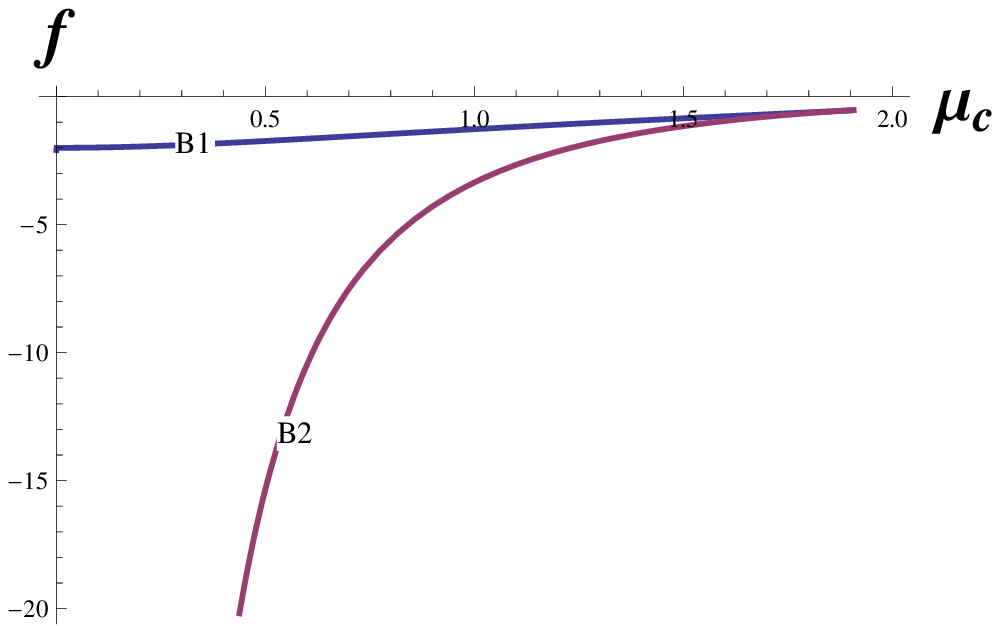, width =2in,height=2in}
\caption{\small{ Here the phase diagram of conical surplus solution is given. The horizontal axes in all the figures is the parameter $\mu_c$. The vertical axes are respectively the spin $3$ charge $\Wc_{CS}$ , stress tensor $\Tc_{CS}$  and free energy $F_{CS}$.}}
\label{CSphasediagram}
\end{figure}

Here as stated above the contractible cycle is spatial and hence we demand that the holonomy of connection defined as $e^{ih}$ where $h=2 \pi(a_z+a_{\bz})$ to be trivial along the contractible cycle. It follows the same holonomy equation as that given in (\ref{holonomy condition}). This choice of center is the same as that for thermal AdS. 
\\\indent The boundary terms that we use (given by the equation above the Drinfeld-Sokolov connection in (\ref{DScondition})) is suited for a study in grand canonical ensemble where, the chemical potentials and temperature are the parameters of the theory. 
\\\indent The first among the two holonomy equations in (\ref{holonomy condition}) can be used to get $\Wc$ in terms of $\Lc$ 
\begin{equation}\label{WCS}
\Wc_{CS} = \frac{1}{12 m \pi T}+\frac{2 \Lc}{3 m T}+\frac{16}{9}m \pi \Lc^2 T 
\end{equation}
\newline and using the second equation we get an equation for $\Lc$ in terms of m and T given by
\begin{equation}\label{CSeq}
-\frac{1}{m T}-\frac{8 \pi \Lc}{m T}+\frac{m T}{3}-\frac{8}{3} m \pi \Lc T+64 m \pi^2 \Lc^2 T+\frac{128}{9} m^3 \pi^2 \Lc^2 T^3-\frac{512}{3} m^3 \pi^3 \Lc^3 T^3+\frac{4096}{27} m^5 \pi^4 \Lc^4 T^5 = 0.
 \end{equation}
 \\ From equation (\ref{stressall}) the stress energy tensor is given by 
\begin{equation}\label{stressCS}
\Tc_{CS}=8 \pi \Lc - 12 m \pi \Wc T - \frac{64}{3} m^2 \pi^2 \Lc^2 T^2,
\end{equation}
 and the spin 3 charge which was $\Wc$ in absence of chemical potential is left unchanged in presence of chemical potential. The free energy given in equation (\ref{freeall}) in this case becomes
 \begin{equation}\label{freeCS}
  F_{CS}= 16 \pi \Lc -8 m \pi\Wc T - \frac{128}{3} m^2 \pi^2 \Lc^2 T^2.
 \end{equation}
% \\\indent The free energy is a function of the temperature and chemical potential and entropy is the Legendre transform of $-\beta$F with the parameters being the charges. The above definition of free energy can be checked to be correct from the fact that the entropy obtained from this is zero \cite{Li:2013rsa} which is what we would expect for a system with no ``horizon''. 
\\\indent Now of the 4 solutions to (\ref{CSeq}), only 2 are real solutions. Using this we can get the solutions for $\Tc_{CS}$ and $F_{CS}$ in terms of m and T. From equation (\ref{CSeq}) we see that all relevant quantities are functions of $\mu_c=mT$. So, the quantities that we use to plot the phase diagrams are $q3(\mu_c)\equiv \Wc(m,T)$, $t(\mu_c)=\Tc_{CS}(m,T)$ and $f(\mu_c)=F_{CS}(m,T)$ in figure (\ref{CSphasediagram}).
\\\indent From the figure (\ref{CSphasediagram}) we see that the blue branch is the branch that goes to thermal AdS (without spin $3$ charge) when $\mu_c\rightarrow 0$. The other branch in red is a special branch where as $\mu_c\rightarrow 0$ we have $\frac{\Wc^\frac{2}{3}}{\Lc}=-\frac{1}{6(2\pi)^\frac{2}{3}}$. This special branch starts from an ``extremal point'' analogous to black holes discussed in ~\cite{Gutperle:2011kf} and \cite{David:2012iu}. Let us call it, the ``extremal branch''. This is a bit of a misnomer as for thermal AdS in any gauge there is no concept of horizon. The two branches merge at the value of the parameter $\mu_c=\frac{3}{4}\sqrt{3+2\sqrt{3}}$, and after that the conical surplus solution ceases to exist. The ``extremal AdS'' branch has an energy which is unbounded from below, but we obtained this branch as a solution to the holonomy conditions which encode the smoothness of the solution. This branch also has the lower free energy of the two branches for all values of the 
chemical potential and temperature. So, unless we have a physical principle (which we are not aware of) the ``extremal AdS'' branch is the most favoured solution and there are no stable solutions to the conical surplus like geometry in presence of chemical potential for spin $3$.

%----------------------------------------------------------------------------------------------------------------------------

\subsection{The black hole solution}
The black hole solution is obtained by demanding that the time circle is contractible and holonomy defined in equation (\ref{timeholonomy}) satisfy the equations in (\ref{holonomy condition}). The holonomy equations in this case are
\begin{eqnarray}\label{hol1}
 2-\frac{32 m^2 \Lc^2}{3}-\frac{4\Lc}{\pi T^2}-\frac{6 m \Wc}{\pi T}&=&0,\nonumber\\
-\frac{128}{9}m^3 \Lc^3+\frac{6 m^3 \Wc^2}{\pi}+\frac{3 \Wc}{2\pi^2 T^3}+\frac{16 m \Lc^2}{\pi T^2}+\frac{12 m^2 \Lc \Wc}{\pi T}&=&0.
\end{eqnarray}
\\\indent Using the same procedure as after (\ref{WCS}) we get the final holonomy equation for the black hole as
\begin{equation}\label{BHeq}
\frac{4 m \Lc}{3}-\frac{64}{3}m^3\Lc^3-\frac{\Lc}{m\pi^2 T^4}+\frac{1}{2 m \pi T^2}+\frac{8 m \Lc^2}{\pi T^2}+\frac{2}{3} m \pi T^2-\frac{64}{9}m^3 \pi \Lc^2 T^2+\frac{512}{27}m^5 \pi \Lc^4 T^2=0.
\end{equation}
\\\indent The free energy of equation (\ref{freeall}) in this case is given by 
\begin{equation}\label{freeBH}
F_{BH}= -16\pi \Lc-8 m \pi \Wc T+\frac{128}{3} m^2 \pi^2 \Lc^2 T^2.
\end{equation}

\begin{figure}[ht]
\epsfig{file=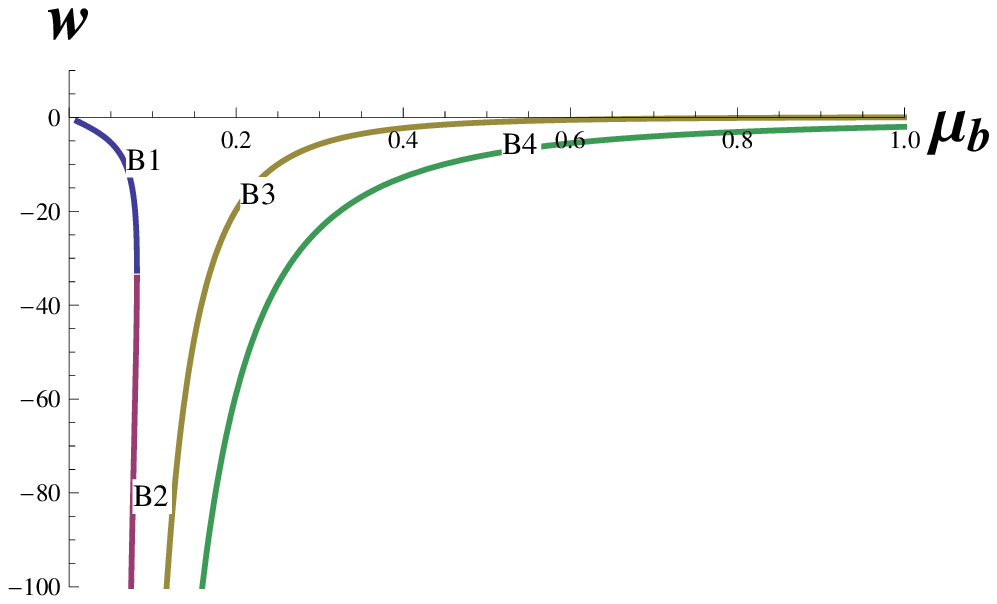, width =3in}\hspace{.5in}
\epsfig{file=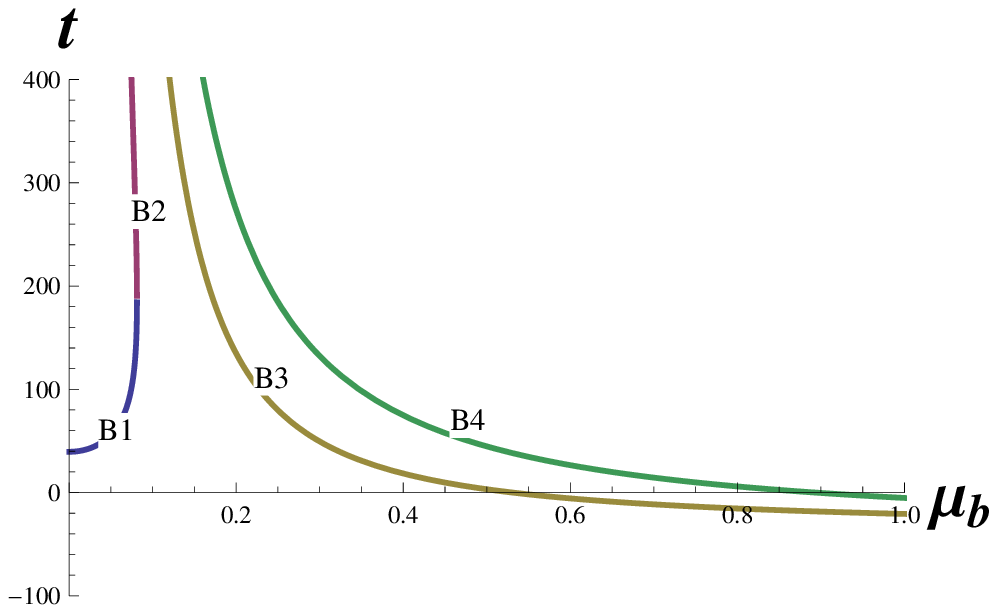, width =3in}\\\\\\
\vspace{0.4in}
\epsfig{file=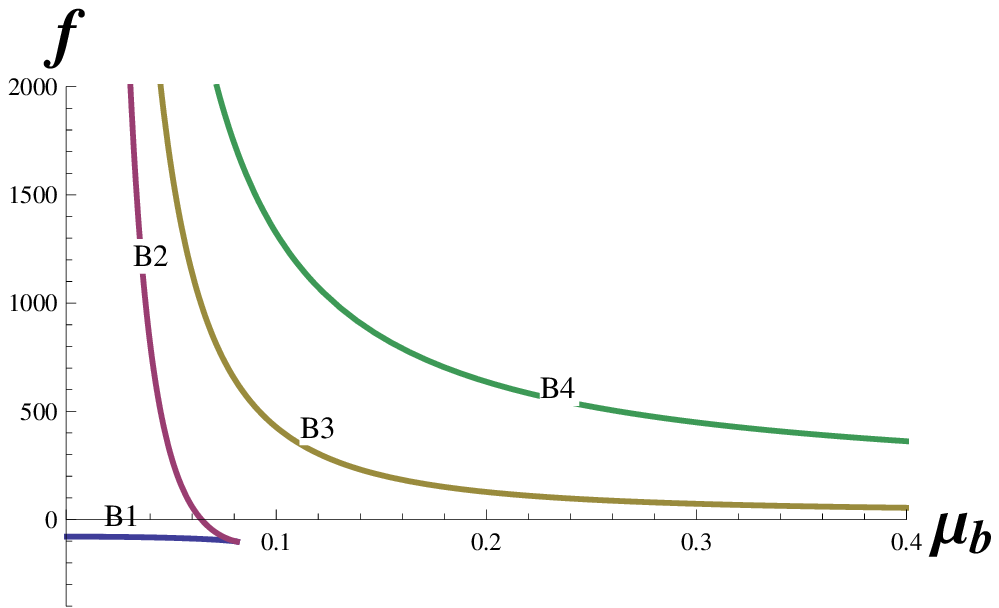, width =3in}\hspace{.5in}
\epsfig{file=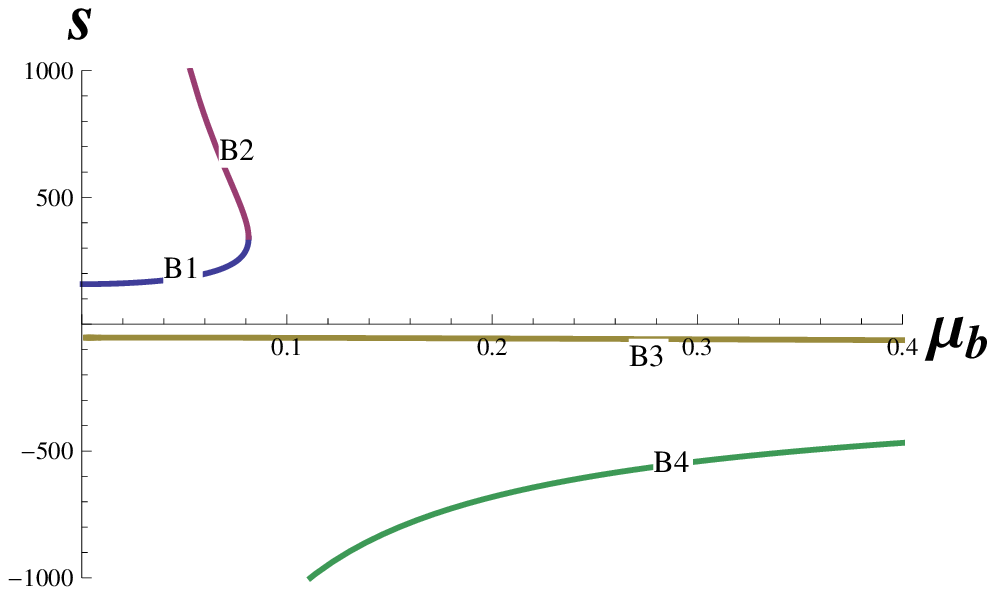, width =3in}
\caption{\small{This figure gives the phase structure for spin 3 black hole. The horizontal axis is $\mu_b$ and the vertical axis on the upper panel are respectively the spin $3$ charge $\Wc_{BH}$ and stress tensor $\Tc_{BH}$ and in the lower panel are free energy $F_{BH}$ and entropy S respectively }}
\label{BHphasediagram}
\end{figure}

The entropy defined in equation (\ref{entropy}) in our case becomes 
\begin{equation}\label{BHentropy}
 S = \frac{32 \pi \Lc}{T}-\frac{256}{3}m^2 \pi^2 \Lc^2 T.
\end{equation}
\\\indent We see that equation (\ref{BHeq}) is an equation for $l=\frac{\Lc}{T^2}$ in terms of $\mu_b=m T^2$. So, $\mu_b$ is a good variable to study the phase structure for the black hole\footnote{The variables which will be used to study the phase structure in terms of $\mu_b$ are
\begin{equation}
 t=\frac{\Tc}{T^2}, \quad w=\frac{\Wc}{T^3}, \quad f=\frac{F}{T^2}, \quad s=\frac{S}{T}.
\end{equation}}
The phase diagram for spin $3$ black hole is given in figure (\ref{BHphasediagram}). We denote the 4 branches of solutions with the following color code- branch-$1$-\textcolor{blue}{Blue},branch-$2$-\textcolor{red}{Red},branch-$3$-\textcolor{YellowOrange}{Orange} and branch-$4$-\textcolor{green}{green}.
\\\indent From the plots we see that branches $3$ and $4$ are unphysical with negative entropy. Branches $1$ and $2$  merge at the point $\mu_b=\frac{3\sqrt{-3+2\sqrt{3}}}{8\pi}$. Beyond this point the black hole solutions cease to exist. For branch $2$ the stress tensor decreases with $\mu_d=m T^2$, i.e. it decreases with $T^2$ if we keep chemical potential m fixed, so this branch has negative specific heat and hence is unstable. So, the branches $1$ and $2$ are in one-to-one correspondence with the large (stable) and small (unstable) black hole solutions in AdS space \cite{Hawking:1982dh, David:2012iu}. For branch $2$, in the limit $\mu\rightarrow0$ we get $\frac{w^\frac{2}{3}}{t}=\frac{1}{6}(\frac{-1}{2\pi})^\frac{2}{3}$, so the branch $2$ evolves from the extremal point having a non trivial configuration at $T=0$. The branches $3$ and $4$ also evolve from the extremal point, but they evolve to unphysical branches. From the free energy plot we see that the BTZ black hole branch 
is the dominant solution 
in the temperature regime 
where it exists for geometries with contractible temporal cycles. %{\color{red} What about the phase transition ?}
%Another interesting thing that we can see is that though the $2nd$ branch is unstable it dominates over the free energy of 1st branch for higher temperatures. We don't know the significance of this yet.

%-----------------------------------------------------------------------------------------------------------------------------------

\subsection{The ``Hawking-Page'' Transition}
\begin{figure}[htp]
\begin{center}
\epsfig{file=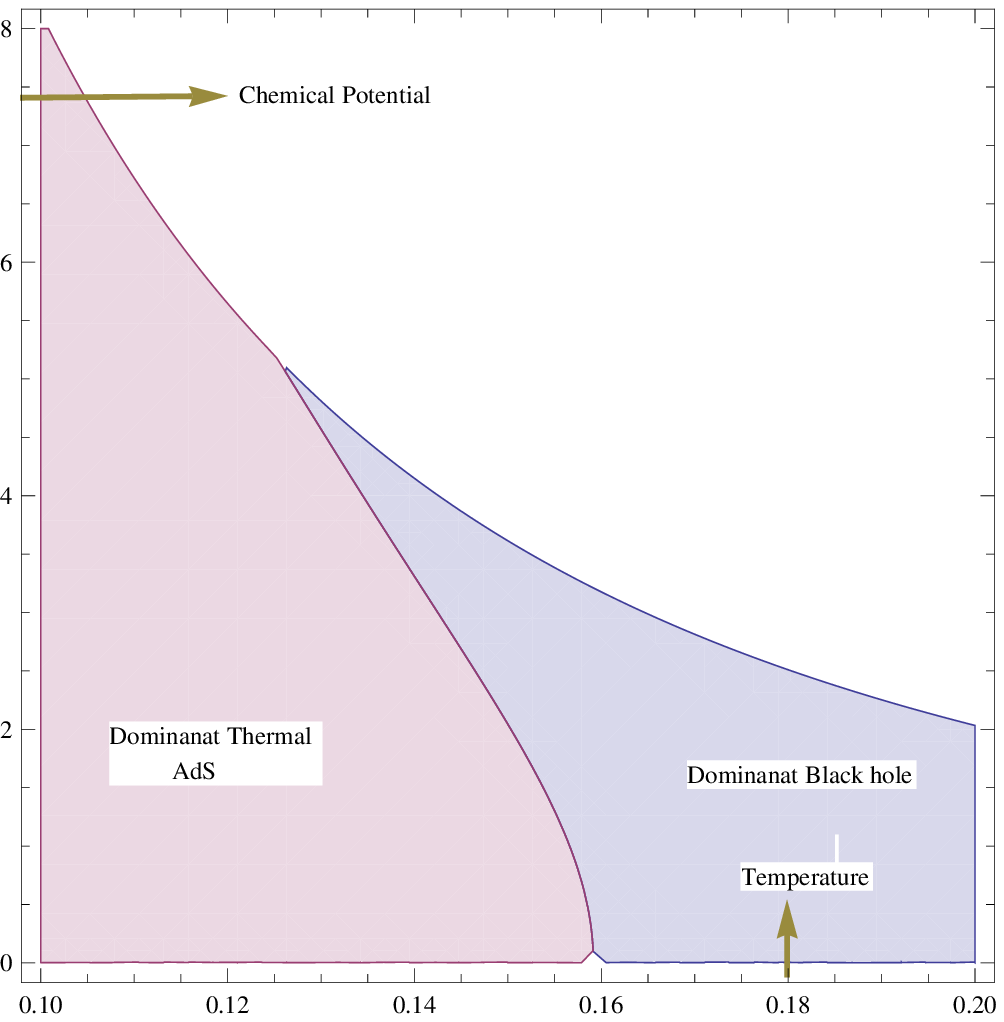, width =3in}\hspace{1in}
\epsfig{file=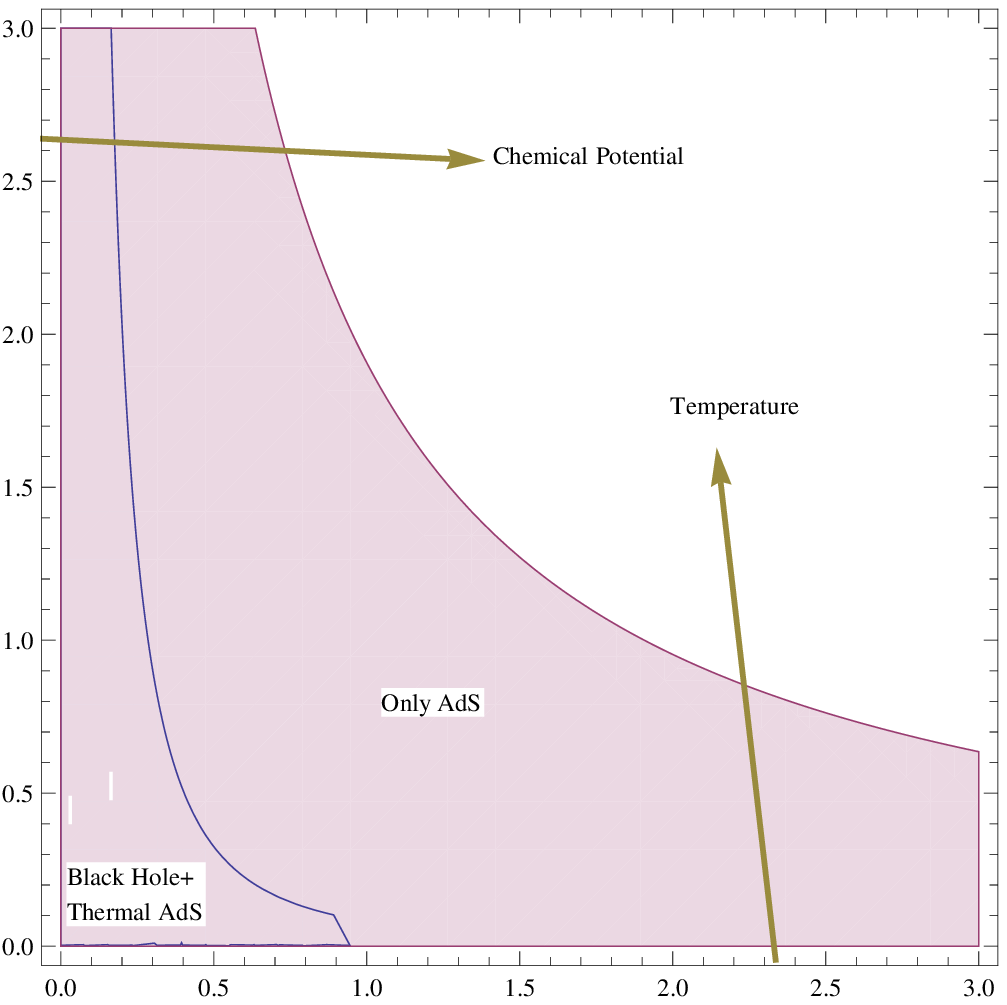, width =3in}
\end{center}
 \caption{\small{In these figures the x axis represents temperature(T) and y axis the chemical potential(m). For the upper figure pink region is where the conical surplus dominates and the blue region indicates where black hole dominates. The boundary between this two regions represents the temperature where the ``Hawking-Page'' transition takes place for a particular chemical potential. The lower figure represents the region of existence of conical surplus and black holes solutions. The black hole solutions exist in the region bound by the axes and the blue line boundary and the conical surplus solution exists in the full coloured region}} 
 \label{hawkingpage}
 \end{figure}

We will now study a phase transition first studied for Einstein-Hilbert gravity with negative cosmological constant on $AdS_4$ in \cite{Hawking:1982dh}. There it was shown that in asymptotically AdS space, out of the two phases $1)$ a gas of gravitons and $2)$ a black hole, the former dominates at low temperature and after a particular temperature the black hole solution becomes more dominant. The dominant phase was obtained by identifying which solution had the lowest free energy for a particular temperature. Both pure AdS (gas of gravitons) and black hole were put at the same temperature by keeping the identification of the time circle at the same value. The free energy was calculated by calculating the on shell action in Euclidean signature with proper added boundary terms. For the $AdS_3$ case, the thermal AdS and BTZ black hole configurations are related by a modular transformation $\t_{BTZ}=-\frac{1}{\t_{AdS}}$. At the point of Hawking-Page transition i.e. $\t_{BTZ}=\t_{AdS}$, $T=\frac{1}
{2\pi}$ (putting the AdS radius to unity). We are 
studying the phase structure in a grand canonical ensemble and we will try to find out the regions in  parameter space where this phase transition takes place. 
\begin{figure}
 \begin{center}
  \includegraphics{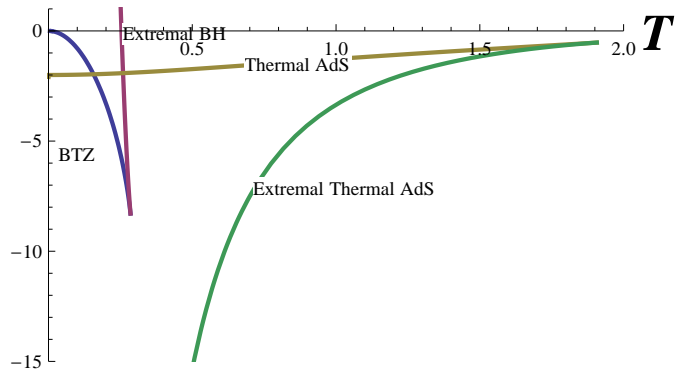}
 \end{center}
\caption{\small{Comparison between free energy of black hole and conical surplus at m=1. The blue branch is the BTZ branch of black hole and red branch is extremal branch. The brown branch is the conical surplus branch which goes to pure AdS in absence of chemical potential and the green branch is the new ``extremal branch'' of conical surplus}}
\label{freeenrgycomparison}
\end{figure}
\\\indent At $m=0$ the temperature at which transition takes place is $T=\frac{1}{2\pi}$. Let us introduce a chemical potential for spin $3$ and see how the temperature deviates from this point. Let us for the moment make an assumption that there is a physical principle behind the "extremal AdS" solution being invalid, so that the thermodynamics is still dominated by the thermal AdS like solution and the black hole solutions.~We will only study the branches which go to BTZ black holes and thermal AdS in the limit $m\rightarrow 0$ i.e. the branch $1$ in both cases. We will assume the following form for the transition temperature after the introduction of a non zero chemical potential m.
\begin{equation}\label{HawkingT}
 T=\frac{1}{2\pi}+\#_1 m+\#_2 m^2+\#_3 m^3+\#_4 m^4+...\quad.
\end{equation}
\\\indent  We shall find the difference between the free energy of the black hole given in equation (\ref{freeBH}) and that of the thermal AdS like solution given in equation (\ref{freeCS}), both at the same temperature and chemical potential. We then find the temperature where this difference is zero which will give the various coefficients in (\ref{HawkingT}) order by order. Upon doing this we arrive at the following temperature where the transition takes place to $O(m^6)$
\begin{equation}\label{seriestemp}
T_{HP}=\frac{1}{2\pi}-\frac{1}{12\pi^3}m^2+\frac{7}{144\pi^5}m^4-\frac{71}{1728\pi^7}m^6+...\quad.
\end{equation}
\\\indent For a chemical potential given by $m=1$ \footnote{This is for the purpose of illustration only as this helps in bringing out all the features nicely in a single diagram. Since the introduction of chemical potential violates the boundary falloff conditions we want $m<<1$ if we want the theory to be studied at high enough temperature as the deformation is by a term of the form $mTW_2e^{2\rho}$ } we plot the free energies of both the black hole and the conical surplus in figure (\ref{freeenrgycomparison}). The color coding is explained in the caption there. We see that the "thermal AdS branch"  dominates over the black hole for low temperature and the BTZ branch black hole solutions take over at higher temperatures. The unstable (extremal) black hole always is the sub-dominant contribution to the free energy compared to BTZ branch. The extremal black hole branch also starts dominating over the thermal AdS as we increase the temperature further. Beyond the 
temperature of existence of the black hole 
the thermal AdS like solutions are the only solutions available. This ``phase transition'' can be explained by a physical argument based on the fact that all even spin fields are self-attractive and all odd spin fields are self-repulsive \footnote{This was brought to our notice by Arnab Rudra and the physical argument arose from a discussion with him.}. So, at very low temperature when there are very few excitations the thermal AdS is the dominating solution. As we increase the temperature the number of excitation of both the spin $2$ and $3$ fields increase but the attractive nature of spin $2$ field dominates and the formation of a black hole is more favourable. Further increasing the temperature causes the number of excitations to increase further and the repulsive nature of spin $3$ dominates over the attractive nature of spin $2$ and makes it unfavourable to form a black hole.
\\ \indent We numerically give the region of dominance of the black hole and thermal AdS like solutions as well as the region of existence of the solutions in figure (\ref{hawkingpage}). We see that the temperature where the "Hawking-Page" transition takes place is lower for higher values of chemical potential. 
 \\\indent From the figure (\ref{hawkingpage}) we see that at any value of chemical potential for high enough temperature, the black hole solution ceases to exist and only thermal AdS like solutions are present. The lower plot in figure (\ref{hawkingpage}) puts this in perspective where we plot the region of existence of the black hole and thermal AdS like solutions . The region of existence of the thermal AdS like solutions is much larger (the full coloured region)than the black hole (region bounded by the axes and the blue line boundary). 
 \\\indent In all this we have to be careful of the fact that introducing a chemical potential corresponds to breaking the asymptotic AdS boundary conditions. The asymptotic AdS falloff conditions  which gives rise to the Virasoro symmetry algebra is $A-A_{AdS}=O(1)$, but by introducing a chemical potential this breaks down to $A-A_{AdS}= m T e^{2\rho}$~\footnote{We have reintroduced the radial dependence by $A=b^{-1}db+b^{-1}ab$}. So,the definition of charges that we are using are not valid if we move too far away from the fixed point. Since, we want to study the property of the system for high enough temperatures we have to confine ourselves to very small values of chemical potential. Also introduction of this deformation induces a RG flow which takes us to another non trivial fixed point in the UV with a completely different spectrum, to be studied next. Hence for large values of m the parameters of the UV fixed point may be the correct parameters to use.
\\\indent  Across the point of transition we see that not only does the stress energy tensor changes sign which is expected, but the spin $3$ charge also changes sign. This can be inferred from the fact that in the allowed regime for conical surplus the spin 3 charge is always positive which can be seen from figure (\ref{CSphasediagram}), and that for black hole it is always negative as can be seen in figure (\ref{BHphasediagram}). 
 %  \begin{figure}
%  \begin{center}
% \includegraphics{hawkingpageRegionplot.eps}
% \end{center}
% \caption{In this figure the x axis represents temperature(T) and y axis the chemical potential(m). The pink region is where the conical surplus dominates and the blue region indicates where black hole dominates. The boundary represents the temperature where the ``Hawking-Page'' transition takes place for a particular chemical potential} 
% \label{hawkingpage}
% \end{figure}

%------------------------------------------------------------------------------------------------------------------------------

\section{The diagonal embedding for $sl(3)$}\label{diagonal}
The definition of $sl(2,R)$ sub-algebra generators in diagonal embedding in terms of generators of principal embedding is $\frac{1}{2}L_0$ and $\pm\frac{1}{4} W_{\pm2}$ as given in~\cite{David:2012iu}~\cite{Ammon:2011nk}. The spectrum here consists of fields of spin $2$, spin $\frac{3}{2}$ and spin $1$. The generators for spin $\frac{3}{2}$ multiplet in the bulk are given by $(W_1,L_{-1})$ and $(W_{-1},L_1)$ and that for spin 1 is $W_0$. The highest weight gauge connection for this theory is given by
\begin{eqnarray}\label{UVconnect}
 a&=&\left(\frac{1}{4}W_2+\Gc L_{-1}+\Jc W_0+\Jc^2 W_{-2}\right)dz+ \frac{\l T_d}{2} \left(L_1+2\Jc L_{-1}-\frac{\Gc}{2}W_{-2}\right)d\bz\nonumber\\
\ba&=&-\frac{\bar{\l}T_d}{2}\left(L_{-1}+2\bar{\Jc}L_1-\frac{1}{2}\bar{\Gc}W_2\right)dz-\left(\frac{1}{4}W_{-2}+\bar{\Gc}L_1+\bar{\Jc}W_0+\bar{\Jc}^2W_2\right)d\bz
\end{eqnarray}
\newline The non rotating condition (\ref{nonrot}) applied here gives 
\begin{equation}
 \bar{\Gc}=-\Gc,\quad \bar{\Jc}=\Jc, \quad \bar{\l}=-\l
\end{equation}
\\\indent Though this embedding looks like an independent theory by itself. But in \cite{Ammon:2011nk} it was shown that after adding a deformation with chemical potential corresponding to spin $\frac{3}{2}$ ($\l$ above), this theory becomes the correct UV behaviour of a theory whose behaviour near IR fixed point is given by the principal embedding studied earlier. If we reintroduce the radial dependence in (\ref{UVconnect}) the leading term comes from $\frac{1}{4}W_2$. So, the way to go to the UV theory from the IR side is to change the coefficient of $W_2$ in $\bz$ component of connection in equation~(\ref{connection}) from $\frac{m T}{2}$ to $\frac{1}{4}$ by a similarity transformation (also found out in \cite{David:2012iu}) 
\begin{eqnarray}\label{simtransform}
 a^{UV}_z=e^{xL_0} a^{IR}_{\bz}e^{-xL_0}&,& \quad a^{UV}_{\bz}=e^{xL_0} a^{IR}_ze^{-xL_0},\quad \nonumber\\
\ba^{UV}_z=e^{-xL_0} \ba^{IR}_{\bz}e^{xL_0}&,& \quad \ba^{UV}_{\bz}=e^{-xL_0} \ba^{IR}_ze^{xL_0}\quad\mbox{where}\quad x= \ln(\sqrt{2 m T}),
\end{eqnarray}
where $a^{UV}$ is the connection given in equation~(\ref{UVconnect}) and $a^{IR}$ is the one given in equation~(\ref{connection}). We see from the map given in (\ref{simtransform}) that the holomorphic and anti holomorphic components change into each other in going from the IR to UV picture. Demanding that equation (\ref{simtransform}) holds we get a relation between parameters of the theories near the UV and IR fixed points like in \cite{David:2012iu} given by
\begin{equation}\label{UVIRparameter}
 \Gc=2\sqrt{2}\pi \Wc (m T)^{\frac{3}{2}},\quad \Jc=-2 \pi \Lc m T, \quad \l T_d=\frac{\sqrt{2}}{\sqrt{m T}}
\end{equation}
The holonomy equation calculated here as in the case of principal embedding is given by
\begin{equation}\label{diagonalholeq}
%-\frac{64 \Jc^3}{3 t^3}+\frac{16 \Jc\pi^2}{3 t}-\frac{512 \Jc^4}{27 t^5 \l^2}+\frac{256 \Jc^2 \pi^2}{9 t^3 \l^2}-\frac{32\pi^4}{3 t \l^2}-\frac{8 \Jc^2 \l^2}{t}-2\pi^2 t \l^2-\Jc t \l^4=0
-\frac{8 \Jc^3}{3\pi^3 T_d^3}+\frac{2\Jc}{3\pi T_d}-\frac{64 \Jc^4}{27 \pi^3 T_d^5 \l^2}+\frac{32 \Jc^2}{9\pi T_d^3\l^2}-\frac{4\pi}{3 T_d \l^2}-\frac{\Jc^2 \l^2}{\pi^3 T_d}-\frac{T_d\l^2}{4\pi}-\frac{\Jc T_d \l^4}{8\pi^3}=0
\end{equation}
The value of spin $\frac{3}{2}$ charge is obtained in terms of the spin $1$ field using the holonomy condition and is given by
\begin{equation}\label{CSspin3by2}
 \Gc_{diag}=-\frac{16\Jc^2}{9 T_d \l}+\frac{4 \pi^2 T_d}{3 \l}+\frac{2 \Jc T_d \l}{3}.
\end{equation}
\\\indent The holonomy equation should also evolve along the RG flow from IR to UV, i.e. the holonomy equation (\ref{diagonalholeq}) should reduce to (\ref{BHeq}), under the transformation of variables given in (\ref{UVIRparameter}) . This happens if over and above the above transformation we assume that the definition of temperature on both limits is the same i.e, $T_d=T$ and the chemical potentials are related by $\l=\frac{\sqrt{2}}{T\sqrt{m T}}$.
\\\indent The definition of the thermodynamic quantities in terms of connection are the same as they were for principal embedding given in \cite{deBoer:2013gz,Li:2013rsa}.  Here their definition in terms of the parameters of diagonal embedding are given by
\begin{eqnarray}\label{diagonal_thermodynamic_variable}
 \Tc_{diag}&=&\frac{16 \Jc^2}{3}-3 \Gc T_d \l+2 \Jc T_d^2 \l^2\nonumber\\
F_{diag}&=&-\frac{32 \Jc^2}{3}+4\Gc T_d \l-4\Jc T_d^2 \l^2\nonumber\\
S_{diag}&=&\frac{64\Jc^2}{3T_d}+8 \Jc T_d \l^2
\end{eqnarray}
\\\indent The equation (\ref{diagonalholeq}) is an equation for $\frac{\Jc}{T_d}$ in $l=\l\sqrt{T_d}$. So, The correct parameter for drawing phase diagram is $l$ and the quantities which are a function of $l$ only, are
\begin{equation}\label{variablechange}
g=\frac{\Gc}{T_d^\frac{3}{2}},\quad j=\frac{\Jc}{T_d},\quad t=\frac{\Tc}{T_d^2},\quad s=\frac{S_{CS}}{T_d},\quad f=\frac{F_{CS}}{T_d^2}
\end{equation}
\begin{figure}[htp]
\epsfig{file=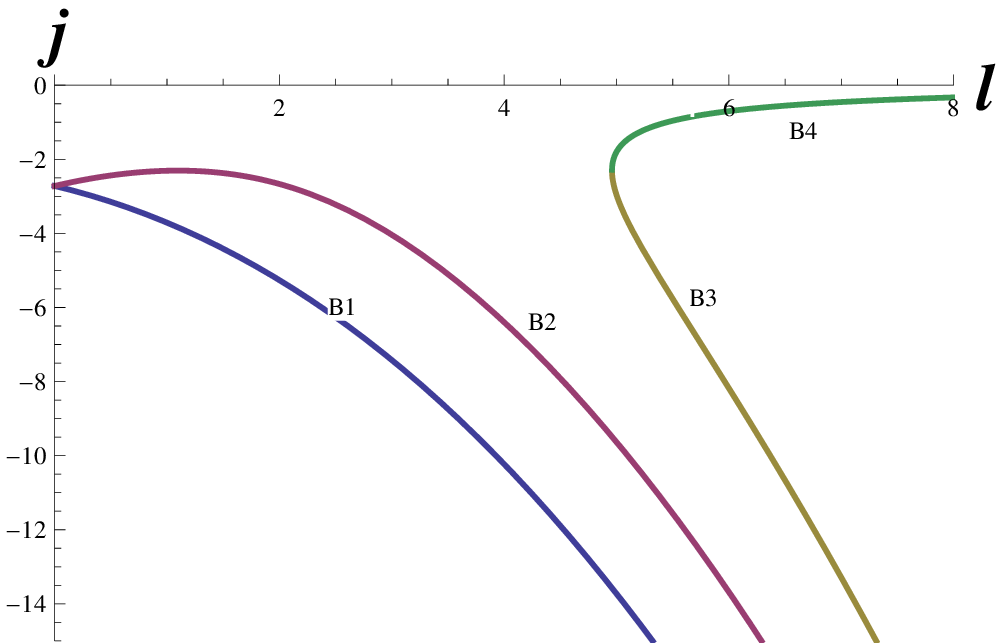, width =3in}\hspace{.1in}
\epsfig{file=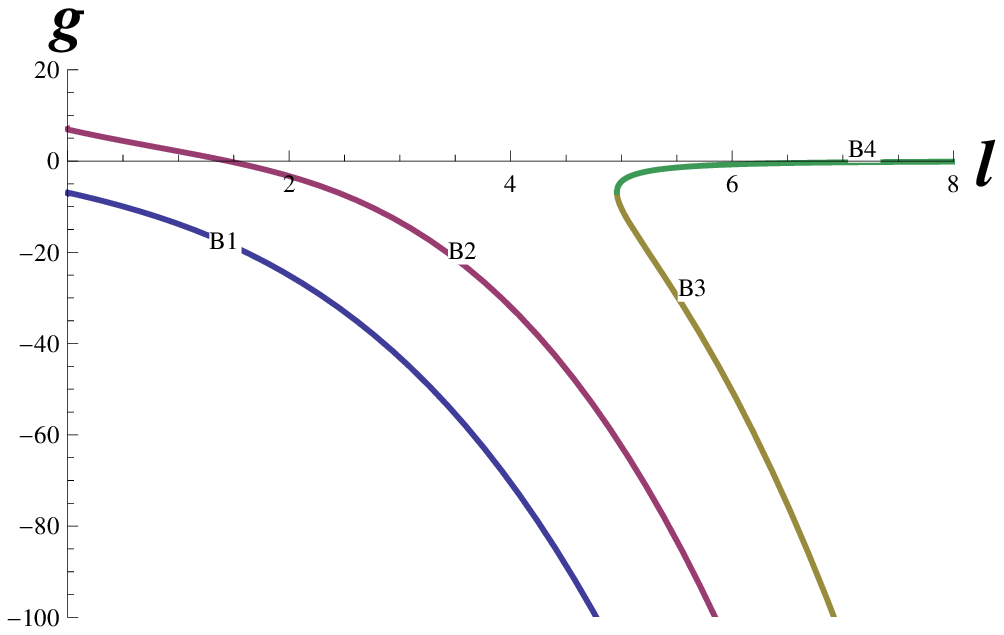, width =3in}\hspace{.1in}
\epsfig{file=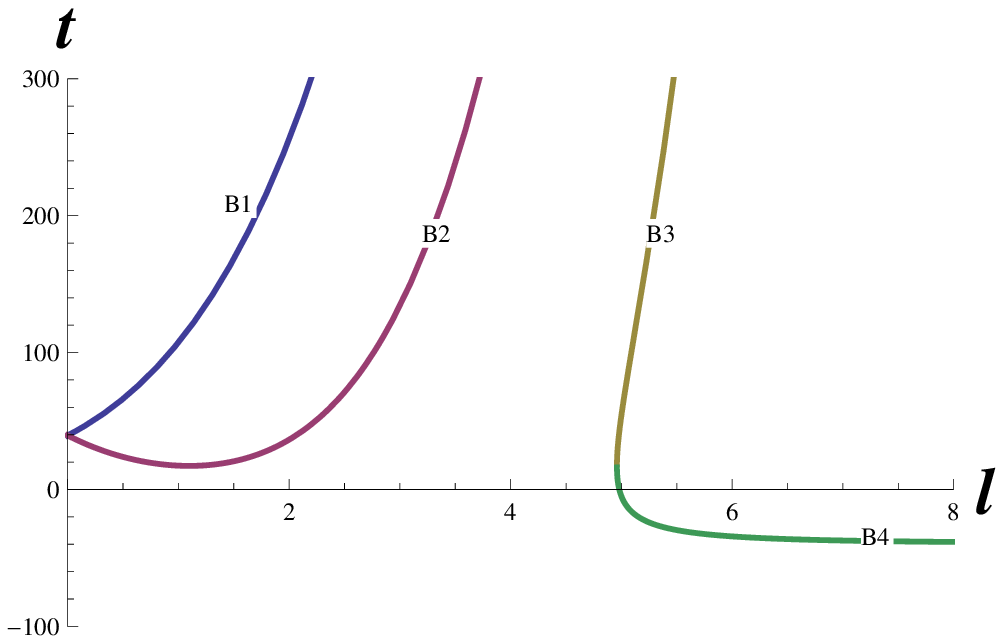, width =3in}
\epsfig{file=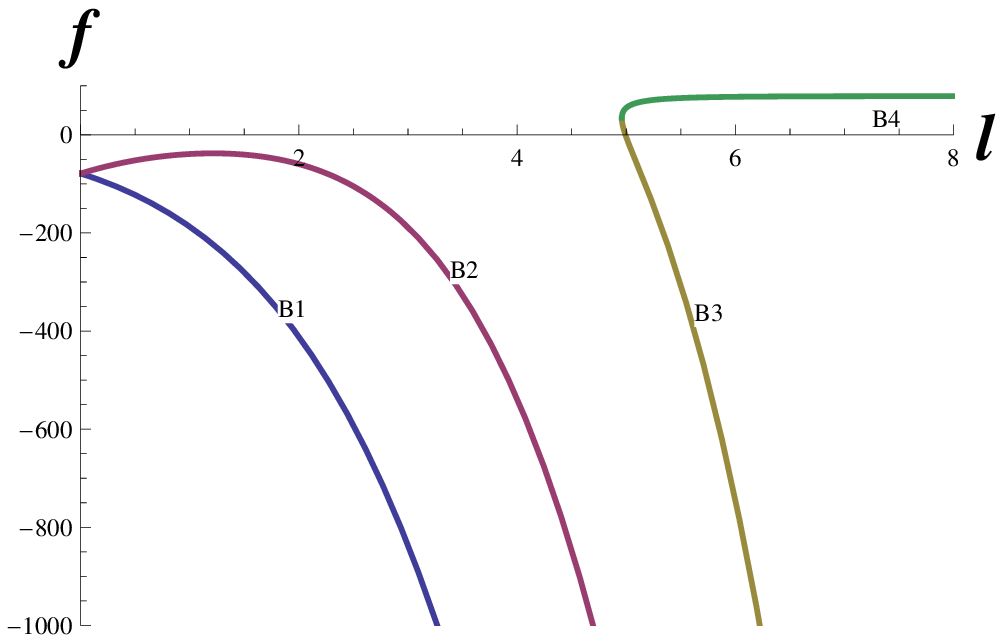, width =3in}\hspace{.5in}
\epsfig{file=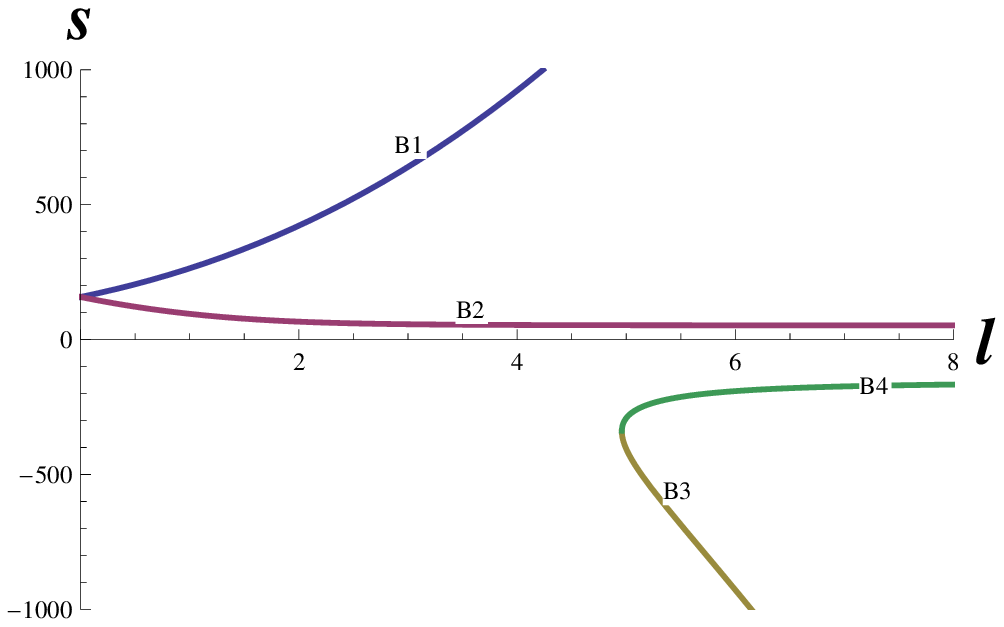, width =3in}
\caption{\small{Phase structure for spin 3 black hole in diagonal embedding. The horizontal axis is $l$ parameter that we used.}}
\label{diagphasediagram}
\end{figure}
\\\indent In the phase diagram for black holes given in figure (\ref{diagphasediagram}) the $4$ branches of solution are color coded as branch $1$-\textcolor{blue}{Blue}, branch $2$-\textcolor{red}{Red}, branch $3$-\textcolor{YellowOrange}{Orange} and branch $4$-\textcolor{green}{Green}. From the phase diagram we see that branches $3$ and $4$ are unphysical because they have negative entropy. If we assume that the chemical potential is fixed at some value then these are plots with respect to square root of temperature. So, if somewhere the gradient of stress tensor is negative then in those region it decreases with temperature and hence the system has negative specific heat. The $2$nd branch has a region of negative specific heat for lower temperature but at higher temperatures it is stable for a given chemical potential. The branch 1 is the dominant solution when we look at the free energy plot. Another interesting thing that we notice is that for branch $1$ and $2$ as $\l\rightarrow0$ we have
\begin{equation}
 \Gc\rightarrow-\frac{2\sqrt{2}\pi^{\frac{3}{2}}T_d^{\frac{3}{2}}}{3^{\frac{3}{4}}},\quad\Jc\rightarrow-\frac{1}{2}\sqrt{3}\pi T_d.
\end{equation}
So, both the spin $\frac{3}{2}$ and spin $1$ charges are non-zero even when the chemical potential corresponding to that charge is zero, i.e. even when the theory is undeformed. This was also obtained in \cite{David:2012iu}. This is different than the principal embedding case where spin $3$ charge goes to zero when the chemical potential goes to zero for the dominant branch. This stems from the fact that $\l\rightarrow0$ limit corresponds to $m\rightarrow\infty$ limit and hence it is not exactly an undeformed theory that we are studying but a theory which has been deformed in IR.
\\\indent We see that this embedding has a valid high temperature behaviour i.e. as $T_d\rightarrow\infty$ we have
\begin{equation}
 \Gc\rightarrow -\frac{-\l^3T_d^3}{2}\rightarrow\frac{l^3 T_d^{\frac{3}{2}}}{2},\quad \Jc\rightarrow-\frac{2\l^2T_d^2}{8}\rightarrow-\frac{2 l^2 T_d}{8}.
\end{equation}
We see that in the high temperature limit the charges have the correct scaling behaviour in terms of the only dimensionful parameter (T) \footnote{l is dimensionless as in terms of l all thermodynamic parameters have correct scaling behaviour with temperature as seen from equation (\ref{variablechange})} and they are real. So, to study the high temperature behaviour the diagonal embedding is the correct theory to use. 
\\\indent In equation (\ref{UVIRparameter}) we have the map between the parameters of the UV and IR theory. Upon substituting the solutions of branch $1,2,3,4$ of the IR theory in this map, it matches respectively with branches $4,3,2,1$ of the UV theory. So, this suggests that along the flow the good solutions in one end go to the bad solutions in the other and vice versa. This is easy to see if we plot branches of $-2 \pi \Lc m T$ and the corresponding branches of $\Jc$ with its parameters $\l$ and $T_d$ replaced by m and T using equation (\ref{UVIRparameter}), the two plots merge with the mentioned identifications of the branches at the two ends. This is expected since from the expression of entropy given in (\ref{entropy}) we see that the sign of the expressions changes if we replace the z-component of connection by $\bz$ component. Also, since we expect the RG flow from IR to UV to happen when m goes from $0$ to $\infty$, we 
see that initially the $\bz$ component of 
the connection acts like a perturbation near IR fixed point as can be seen from equation (\ref{connection}), but near UV fixed point due to $m\rightarrow\infty$ $\bz$ component is dominant part. So, the sign of entropy of the branches changes between the UV and IR fixed points and hence the good and the bad solutions get swapped. This is the reason for the bad branches in IR being able to explain the high temperature behaviour of the theory in the UV.

%------------------------------------------------------------------------------------------------------------------------

\section{Summary and Discussions}\label{summary}
\subsection{Comparison with earlier works}
At first glance our analysis may look very similar to \cite{David:2012iu}. But we differ from \cite{David:2012iu} in the following respect
\begin{itemize}
\item \textbf{Convention:}We use a particular set of boundary terms in our bulk action given in \cite{deBoer:2013gz} so that the variation of the full action is like $\delta I \sim \Tc\delta\tau+Q_i\delta m^i$ which ensures that our on shell partition function is of the form $Z=e^{\t\Tc+m^iQ_i}$. But in \cite{David:2012iu} they added a boundary term which made sure that the variation of the on shell action is of the form $\delta I \sim \Lc\delta\tau+Q_i\delta(\tau m^i)$, so that in that case the on shell partition function is like $Z=e^{\t\Lc+\t m^iQ_i}$, with the convention for connection being that of (\ref{connection}). In other words our connection are suited to a convention where the chemical potential corresponding to the spin $3$ charge is given by 'm' whereas the convention used in \cite{David:2012iu} is suited to the chemical potential being proportional to '$\tau$m'. Due to this difference in convention our physical quantities are not finite in the $m\rightarrow 0$ or $T\rightarrow 0$, whereas they 
are finite for the conventions used in \cite{David:2012iu}. For a better comparison of our results with that of \cite{David:2012iu} we need to add a boundary term to our action so that its variation becomes $\delta I \sim \Tc\delta\tau+Q_i\delta (m^iT)$, which we leave for future investigation. 
 
 %In the study of black hole phase structure in principal embedding in \cite{David:2012iu} they get one unphysical branch and $3$ physical branch. Of the remaining $3$ branches they have one unstable %branch with negative specific heat. Of the remaining 2 stable branches they have one branch with the expected scaling behavior and one with wrong scaling behaviour with temperature for quantities having %%a CFT description. In our case we get $2$ unphysical and $2$ physical branch. Our BTZ branch has the correct scaling behavior $(e.g. Stress \; Tensor \propto T^2)$ whereas our extremal branch apparently %%does not have one $(\propto \frac{1}{m^2 T^2})$. But using the correct dimensionless thermodynamical variable that we use to plot our phase plots, the scaling of the extremal branch goes like $\propto %%\frac{T^2}{\mu^2}$ and hence, it too has the correct scaling behaviour. So, both our branches can have a consistent CFT description. 
\item\textbf{Qualitative difference in phase structures:} We will now detail the qualitative difference between the phase structure in principal embedding obtained by us in canonical formalism and obtained in \cite{David:2012iu} using holomorphic formalism.
\begin{itemize}
\item In canonical formalism we obtain two physical and two unphysical (due to negative entropy) branches but in holomorphic formalism there are 3 physical and one unphysical branches.
\item In holomorphic formalism out of the three physical branches only one (the BTZ branch) had the correct scaling behaviour with temperature for the thermodynamic quantities at low temperatures. In our case we showed that all branches have the correct temperature scaling behaviour when written in terms of correct dimensionless variables. So, only two of the branches in \cite{David:2012iu} can have a CFT like interpretation whereas in our case all branches have a correct CFT like interpretation but two of those branches are unphysical only due to the fact that they have negative entropy. 
\end{itemize}
\item %In \cite{David:2012iu} they argued that the third real and physical branch which had the wrong scaling behaviour with temperature at high temperatures infact had the correct scaling behavior when looked at from the diagonal embedding. They argued this by stating that in the diagonal embedding, of the two real branches near the fixed point, the one with the lower free energy must map to the lower free energy branch among the two surviving branches from the principal embedding \footnote{There this was the branch with positive entropy for principal embedding}. Whereas in our case we explicitly show that the good branches in the principal embedding map to the bad branches in diagonal embedding and vice-versa by giving an explicit one-to-one mapping between the branches at the two ends. We also show that in terms of the temperature and the dimensionless parameter used the thermodynamic quantities have the correct scaling 
%behaviour at high temperature in the diagonal embedding. 
\textbf{Method of resolution of the high temperature behaviour by invoking diagonal embedding:} In \cite{David:2012iu} this was done by stating that the the third branch in the low temperature regime actually maps to one of the two real solutions in the high temperature regime (the one with the lower free energy; the one with the higher free energy is thought of as having its low temperature behaviour defined by the higher free energy carrying unphysical branch in principal embedding). They showed that the thermodynamic quantities in this branch when written in terms of diagonal embedding variables have the correct scaling behaviour with temperature and hence is the correct high temperature description of a system with $SL(3,R)\times SL(3,R)$ gauge group.
\\In our work we suggested a slightly different mechanism of how the diagonal embedding comes to the rescue, giving the system a meaningful description in the high temperature regime. We showed that for any chemical potential the two physical branches of the principal embedding only survive upto a particular temperature. But there are also two unphysical branches which survive at all temperatures though they have negative entropy. Similarly, for diagonal embedding there are two unphysical and two physical branches of solution and all of them survive at high temperatures. We then used the map between the parameters of the theory in the principal and diagonal embedding given in (\ref{UVIRparameter}) to show that the thermodynamic quantities for the unphysical branches in principal embedding map to the thermodynamic quantities of the physical branches in diagonal embedding and vice versa. We then argued that the the unphysical branches of the principal embedding which survive at all temperatures actually become 
the physical branches of the diagonal embedding. This explains the high temperature behaviour of the system with $SL(3,R)\times SL(3,R)$ gauge group. Since, we have no physical branch in the principal embedding which survives for very high temperatures, we have no other option but to use the diagonal embedding as the correct picture at very high temperatures.
\item \textbf{Additional new features mentioned by us:} a) In addition to this we studied the thermodynamics of the thermal AdS like solutions in the principal embedding and in the process we were able to show that a `` Hawking-Page'' like transition takes place in the low temperature regime. Also, after a certain temperature when the black hole solutions in the principal embedding cease to exist the thermal AdS like solution again takes over as for a particular chemical potential its regime of existence extends to a higher temperature than that of the black hole. As we have stated earlier this phenomenon happens most probably due to the self repulsive nature of spin $3$ fields due to which at high enough temperatures black hole formation is prevented. 
\\b) We also found out the existence of a second problematic branch in the phase structure for the thermal AdS like solution in presence of chemical potential. This we called the "extremal AdS" branch. At first glance it is thermodynamically the most stable branch of the two conical surplus like solutions and two physical black hole like solutions. But this branch has the pathological property of having its energy unbounded from below. This branch fits in the criteria of allowed solution in terms of triviality of holonomy along contractible cycles. So, to get rid of this branch we must figure out some other physical condition which in particular is not satisfied by the extremal AdS like solutions. We have not been able to find this new physical condition as yet. In absence of that we must say that this system with gauge group $Sl(3,R)\times SL(3,R)$ do not have thermodynamically stable solutions at low temperatures and that this system defines only black holes with spectrum in diagonal embedding for high 
temperatures. If we manage to come up with a physical criteria to get rid of this pathological solution then the phase structure of the system will be the one mentioned in earlier parts of this section.
\end{itemize}
%\indent Let us now shift our focus back to the ``extremal thermal AdS like branch``. It has the lowest value for free energy at all points in the parameter space where it exists among all solutions. But from the energy plots we see that its energy is unbounded from below for $m T\rightarrow0$. So, this is not a physically acceptable solution. So, this branch will not destroy the overall phase structure as this is an unphysical branch.

\subsection{The phase structure at arbitrary point in $\mu$-T space}
\indent We must mention that we have given the phase structure in regions near $\mu=0 (\lambda=\infty)$ i.e. the IR fixed point and $\lambda=0(\mu=\infty$) i.e. the UV fixed point. It is difficult to obtain the phase structure at arbitrary point in the $\mu-T$ phase space. The reason being that only near these two extremum values of $\mu$ do we have assymptotic AdS regions and definitions of charges which follow a $W_3$ algebra for Poisson brackets. Far from these fixed points we cannot and should not trust the expressions for charges. In \cite{Compere:2013nba} it was shown that for arbitrary values of spin $3$ chemical potential they could go back to connections ( by appropriate coordinate transformation followed by gauge transformations) which are again those of assymptotic AdS in new coordinate system. They however showed that the coordinate transformation used for the above mentioned procedure breaks down the boundary light cone structuure for values of $\mu$ greater than a threshold value. They showed 
that beyond this point the thermodynamic quantities turn out to be complex numbers. These bounds on chemical potentials exactly matches with the bound that we found on chemical potential at which branches in our solution space merge and cease to exist after that (after taking care of different conventions in both works).

\subsection{Comparison with Vasiliev system}
\indent The system that we studied here is very different from Vasiliev system. Our system consists solely of one additional higher spin field on top of gravity in $3$ dimensions, whereas in Vasiliev system there is an infinite tower of higher spin fields along with some matter fields. The phase structure of Vasiliev system in $AdS_4$ was studied in \cite{Shenker:2011zf} by studying the dual system of the singlet sector of O$(N)$ vector model on a $3$ dimensional sphere. They observed that the entropy jumps from O$(1)$ to O$(N^2)$ at a temperature O$(\sqrt{N})$ instead of O$(1)$. So, there are no thermodynamically stable large black holes at temperatures of O$(1)$ in this system. Similar study was undertaken in $W_{\infty}[0]$ CFTs in $2$ dimensions in \cite{Banerjee:2012aj} and they found similar results. This was attributed to the presence of a very huge spectrum of light states in these CFTs which smoothen out the phase transition. 
 \\\indent The system that we studied is much closer to the pure gravity system than it is to the Vasiliev system in its content, so the results that we obtained are much closer to the pure gravity system in the sense that the transition temperature is close to $\frac{1}{2\pi}$ which is the Hawking Page temperature for pure $AdS_3$. Though a unitary CFT dual to these system with finite spin is not known (interesting non-unitary CFT dual to finite spin systems have been worked out in \cite{Gaberdiel:2012ku,Perlmutter:2012ds}) yet but if we extrapolate the results from the above mentioned works, then the transition temperature should be O$(\sqrt{N})\sim$ O$(1)$ which is what we obtain. Secondly these systems by similar argument cannot have a huge spectrum of light states which smoothen out these transitions and hence there is no barrier for these transitions to take place.

%---------------------------------------------------------------------------------------------------------------

\section{Further Directions}\label{future}
%\indent We did our calculations in the canonical formalism which is different from the holomorphic formalism used in \cite{David:2012iu}. But the canonical formalism is in apparent disagreement with the CFT calculations of \cite{Gaberdiel:2012yb}. But in the canonical formalism the thermodynamic quantities are obtained much more naturally so we think that there has to be a way to see if it matches with the CFT calculation in the highest spin going to $\infty$ limit, where the CFT calculations have been done. A possible solution for this was suggested in \cite{Compere:2013nba}.
\indent In the the recent paper \cite{Henneaux:2013dra} where it has been proposed that for higher spin theories the correct way to add chemical potential preserving the Brown-Henneaux fall off conditions necessary for definition of charges that we are using is to add them along the time component of the connection rather than the antiholomorphic component. In the light of this our analysis should be redone to see if some extra features emerge other than what we have already presented. The ideal situation would be to derive the the asymptotic charges in the presence of a chemical potential exactly.  

%------------------------------------------------------------------------------------------------------------------

\section{Acknowledgement}
We would like to thank Rajesh Gopakumar, Ashoke Sen, Anirban Basu, Justin David, Arnab Rudra, Arjun Bagchi, Suvankar Dutta for many helpful discussions. We would also like to thank Souvik Datta for having discussions which cleared many aspects of higher spin black holes and modular transformation to us. A.S. would also like to thank Ushoshi Maitra and Sabyasachi Tarat for general discussions on many aspects of this work. A.S. would also like to thank Joydeep Chakrabortty for help with figures in Mathematica. We would also like thank Rajesh Gopakumar for going through our draft carefully and suggesting many improvements. We would also like to thank the organisers of NSM $2013$ held at Indian Institute of Technology, Kharagpur for hospitality provided while the work was being finished. 

\end{document}